\documentclass{ws-procs9x6}

\usepackage{mathrsfs}

\makeatletter
\def\simleq{\mathrel{\mathpalette\gl@align<}}
\def\simgeq{\mathrel{\mathpalette\gl@align>}}
\def\gl@align#1#2{\lower.6ex\vbox{\baselineskip\z@skip\lineskip\z@
     \ialign{$\m@th#1\hfill##\hfil$\crcr#2\crcr\sim\crcr}}}
\makeatother

\newcommand{\bra}{\langle}
\newcommand{\ket}{\rangle}
\newcommand{\braket}[1]{\bra #1 \ket}
\newcommand{\qq}{\braket{\bar{q}q}}

\newcommand{\qGq}{g\braket{\bar{q}\sigma_{\mu\nu}G_{\mu\nu} q}}

\newcommand{\ucap}[1]{\uppercase{#1}}

{\end{minipage}\par}

\begin{document}

\title{
The Study of Chiral Restoration using\\
the Quark-Gluon Mixed Condensate $\qGq$\\
in Lattice QCD at Finite Temperature%
\footnote{
\ucap{T}he lattice 
\ucap{QCD}
\ucap{M}onte 
\ucap{C}arlo simulations
have been performed on 
\ucap{NEC} 
\ucap{SX-5} at 
\ucap{O}saka 
\ucap{U}niversity and
\ucap{IBM}
\ucap{POWER4}
at 
\ucap{T}okyo 
\ucap{I}nstitute of 
\ucap{T}echnology.
}
}

\author{T. Doi$^1$, N. Ishii$^2$, M. Oka$^1$ and H. Suganuma$^1$}

\address{$^1$
Department of Physics, Tokyo Institute of Technology,\\
Ohokayama 2-12-1, Meguro, Tokyo 152-8551, Japan\\
E-mail: doi@th.phys.titech.ac.jp
}

\address{$^2$
Radiation Laboratory, 
The Institute of Physical and Chemical Research,\\
(RIKEN),
Hirosawa 2-1, Wako, Saitama, 351-0198, Japan}


\maketitle

\abstracts{
\vspace*{-1mm}
The quark-gluon mixed condensate $\qGq$ is studied 
using SU(3)$_c$ lattice QCD with the Kogut-Susskind fermion 
at the quenched level.
Using the lattices as 
$\beta=6.0$ with $16^3\times N_t (N_t=16,12,10,8,6,4)$,
$\beta = 6.1$ with $20^3\times N_t (N_t=20,12,10,8,6)$ and 
$\beta = 6.2$ with $24^3\times N_t (N_t=24,16,12,10,8)$
in high statistics of 100-1000 gauge configurations,
we perform accurate measurement of the thermal effects 
on $\qGq$ as well as $\qq$ in the chiral limit.
We find that the thermal effects on both the condensates
are very weak except for the vicinity of $T_c$, while
both the condensates suddenly vanish around $T_c \simeq 280 {\rm MeV}$,
which indicates strong chiral restoration near $T_c$.
We also find that the ratio 
$m_0^2 \equiv \qGq / \qq$ 
is almost independent of the temperature
even in the very vicinity of $T_c$, which
means that these two different condensates obey the same critical behavior.
This nontrivial similarity between them would impose constraints on 
the chiral structure of the QCD vacuum near $T_c$.
\vspace*{-1mm}
}


\section{Introduction}

Fruitful phenomena in hadron physics, 
such as spontaneous chiral-symmetry breaking 
and color confinement,
originate from the non-perturbative nature 
of QCD, or its nontrivial vacuum structure.
The study of the vacuum structure of QCD is 
especially interesting 
at finite temperature/density,
because it is expected that QCD has various phase 
structures in the phase diagram, and chiral restoration 
and color deconfinement can be observed.
Considering that the on-going RHIC experiments 
are actually tackling this subject,
we here present the theoretical study of 
finite temperature QCD.
In this point,
condensates are useful physical quantities because 
they characterize the nontrivial QCD vacuum.
At finite temperature, 
we can probe the change of the QCD vacuum,
by examining the thermal effects on the condensates.

Among various condensates, we study 
the quark-gluon mixed condensate $\qGq$ for the following reasons.
First, we note that the mixed condensate is 
a chiral order parameter independent of $\qq$,
since it flips the chirality of the quark as
\begin{eqnarray}
\qGq =
  g\braket{\bar{q}_R (\sigma_{\mu\nu}G_{\mu\nu}) q_L}
+ g\braket{\bar{q}_L (\sigma_{\mu\nu}G_{\mu\nu}) q_R}.
\end{eqnarray}
Therefore, $\qGq$, as well as $\qq$, are important indicators
on the chiral structure of the QCD vacuum.
In particular, 
the thermal effects 
on both the condensates and the comparison of them
will reveal mechanism of the
chiral restoration of the QCD vacuum at finite temperature.
Second, in contrast to $\qq$, 
$\qGq$ represents a direct correlation 
between quarks and gluons in the QCD vacuum, and thus
characterizes different aspect of the QCD vacuum. 
%
%
Third,
the mixed condensate is relevant in 
various QCD sum rules, especially in baryons\cite{Dosch},
light-heavy mesons\cite{Dosch2}
and exotic mesons\cite{Latorre}.
%
%
%
In this point, the quantitative estimate of the thermal
effects on $\qGq$ has an impact on hadron phenomenology at finite temperature
through the framework of the QCD sum rule.
Therefore, it is desirable to estimate $\qGq$ at zero/finite temperature
by a direct calculation from QCD, 
such as lattice QCD simulations.

So far, the mixed condensate $\qGq$ at zero temperature has been 
analyzed phenomenologically in the QCD sum rules\cite{Bel}.
In the lattice QCD, 
there has been no result except for a rather preliminary 
but pioneering work\cite{K&S} done long time ago.
Recently, new lattice calculations have been performed 
by us\cite{DOIS:qGq,DOIS:T} using the Kogut-Susskind (KS) fermion,
and by other group\cite{twc:qGq} using the Domain-Wall fermion.
At finite temperature, however, there has been no 
result on $\qGq$ except for 
our early results\cite{DOIS:T}.
Therefore, we here present the extensive results
of the thermal effects on $\qGq$ as well as $\qq$,   
including the analysis near the critical temperature\cite{DOIS:T2}.


\section{The Lattice Formalism}

We calculate the condensates $\qGq$ as well as $\qq$
using SU(3)$_c$ lattice QCD with the KS-fermion 
at the quenched level.
We note that the KS-fermion preserves the explicit chiral symmetry
for the quark mass $m=0$, which is a desirable feature
to study both the condensates of chiral order parameters.

The Monte Carlo simulations are performed with the 
standard Wilson action for $\beta=6.0, 6.1$ and $6.2$.
In order to calculate at various temperatures, 
we use the following lattices as
\vspace*{-1mm}
\begin{enumerate}
\item  $\beta = 6.0$,\ \ $16^3\times N_t\ \ (N_t=16,12,10,8,6,4)$,
\item  $\beta = 6.1$,\ \ $20^3\times N_t\ \ (N_t=20,12,10,8,6)$,
\item  $\beta = 6.2$,\ \ $24^3\times N_t\ \ (N_t=24,16,12,10,8)$.
\end{enumerate}
\vspace*{-1mm}
We generate 100 gauge configurations for each lattice.
However, in the vicinity of the chiral phase transition point,
namely, $20^3\times 8$ at $\beta=6.1$ and $24^3\times 10$ at
$\beta=6.2$, the fluctuations of the condensates get larger.
We hence generate 1000 gauge configurations for the 
above two lattices to improve the estimate.
The lattice units are obtained as
$a\simeq 0.10, 0.09, 0.07 {\rm fm}$ for 
$\beta = 6.0, 6.1, 6.2$, respectively, 
which reproduce 
the string tension $\sigma = 0.89 {\rm GeV/fm}$.
We calculate the flavor-averaged condensates as
\begin{eqnarray}
a^3 \qq
&=& - \frac{1}{4}\sum_f {\rm Tr}\left[ \braket{q^f(x) \bar{q}^f(x)} \right], \\
a^5 \qGq
&=& - \frac{1}{4}\sum_{f,\ \mu,\nu}{\rm Tr}
        \left[ \braket{q^f(x) \bar{q}^f(x)} \sigma_{\mu\nu} G_{\mu\nu}
\right],
\end{eqnarray}
where SU(4)$_f$ quark-spinor fields, $q$ and $\bar{q}$, are converted into 
spinless Grassmann KS-fields $\chi$ and $\bar{\chi}$
and the gauge-link variable.
We adopt the clover-type definition for
the gluon field strength $G_{\mu\nu}$ on the lattice 
in order to eliminate $\mathscr{O}(a)$ discretization error.
The more detailed formula for the condensates and gluon field strength
are given in Ref.~\refcite{DOIS:qGq}.

In the calculation of the condensates,
we use the current-quark mass $m = 21,36,52 {\rm MeV}$.
For the fields $\chi$, $\bar{\chi}$, the anti-periodic condition 
is imposed at the boundary in the all directions.
We measure the condensates 
on 16 different physical space-time points of $x$ ($\beta=6.0$) or 
2 points of $x$ ($\beta = 6.1,6.2$)
in each configuration.
Therefore, at each $m$ and temperature, 
we achieve high statistics in 
total as
1600 data ($\beta=6.0$) or 200 data ($\beta=6.1,6.2$).
Note that in the vicinity of the critical temperature, $20^3\times 8$ 
($\beta =6.1$)
and $24^3\times 10$ ($\beta =6.2$), 
we take up to 2000 data so as to achieve  high statistics 
and to guarantee reliability of the results.


\section{The Lattice Results and Discussions}

We calculate the condensates $\qGq$ as well as $\qq$
at each quark mass and temperature.
We observe that both the condensates show 
a clear linear behavior against the quark mass $m$,
which can be typically seen in Ref.~\refcite{DOIS:qGq}.
We therefore fit the data with a linear function and determine 
the condensates in the chiral limit.
Estimated statistical errors are typically less than 5\%,
while the finite volume artifact is estimated to be about 1\%, 
from the check on the dependence of the condensates on
boundary conditions\cite{DOIS:qGq}.

\begin{figure}[htb]
\hfill
\begin{center}
 \begin{minipage}{55mm}
 \includegraphics[scale=0.22]
 {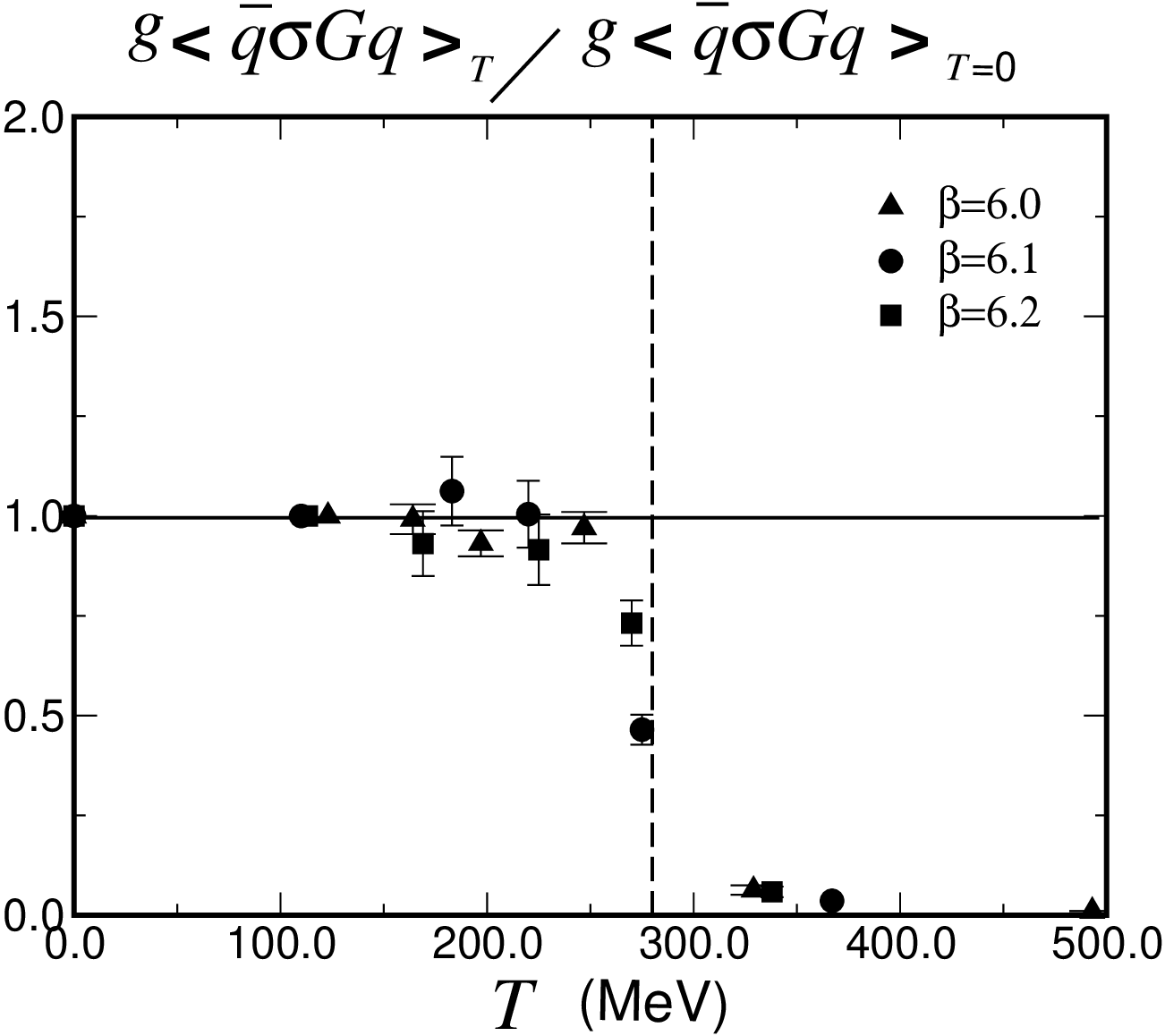}
 \caption{
 Thermal effects on the mixed condensate $\qGq$ 
 v.s. the temperature $T$.
 The vertical dashed line denotes the critical 
 temperature $T_c \simeq 280 {\rm MeV}$
 at the quenched level.
 }
 \label{fig:qGq_finite_T}
 \end{minipage}
\hfill
 \begin{minipage}{55mm}
 \begin{center}
 \vspace*{-10mm}
 \includegraphics[scale=0.22]
 {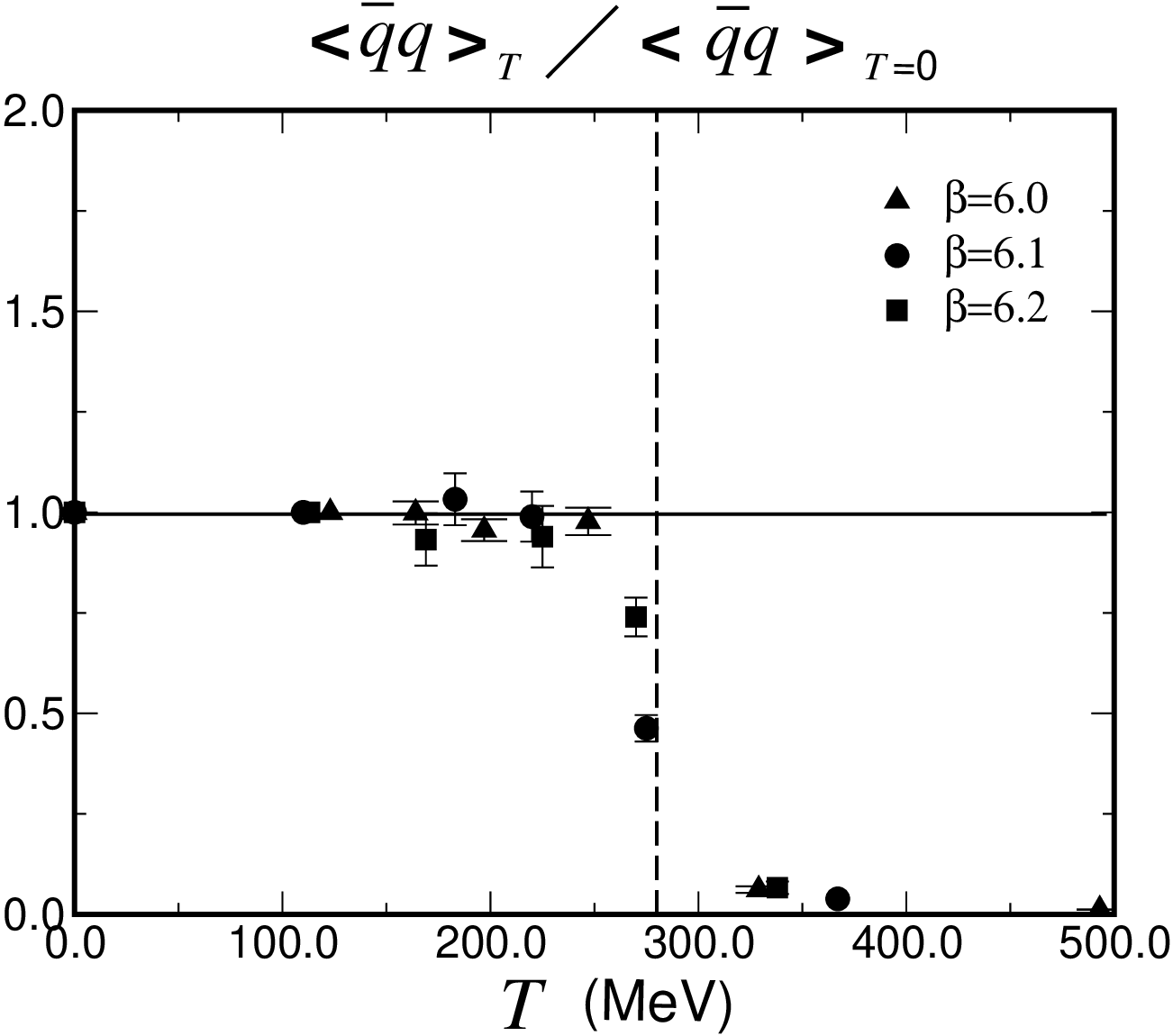}
 \caption{
 Thermal effects on the quark condensate $\qq$ v.s. $T$.
 }
\label{fig:qq_finite_T}
 \end{center}
 \end{minipage}
\end{center}
\end{figure}

For each condensate, 
we estimate the thermal effects 
by taking the ratio between the values at finite 
and zero temperatures.
The renormalization constants are expected to be
canceled in this ratio 
because there is no operator mixing for 
both the condensates
in the chiral limit\cite{Narison2}.
In figure~\ref{fig:qGq_finite_T}, we plot the thermal effects on 
$\qGq$.
We find a drastic change of $\qGq$ around the critical temperature
$T_c \simeq 280 {\rm MeV}$. This is a first observation of 
chiral-symmetry restoration  in terms of $\qGq$.
We also find that 
the thermal effects on $\qGq$ are very weak
below the critical temperature, 
namely, $T \simleq 0.9 T_c$.
In figure~\ref{fig:qq_finite_T}, 
the same features can be also seen for $\qq$.

\vspace*{-5mm}

\vspace*{-5mm}

\begin{figure}[htb]
\begin{center}
\includegraphics[scale=0.24]
{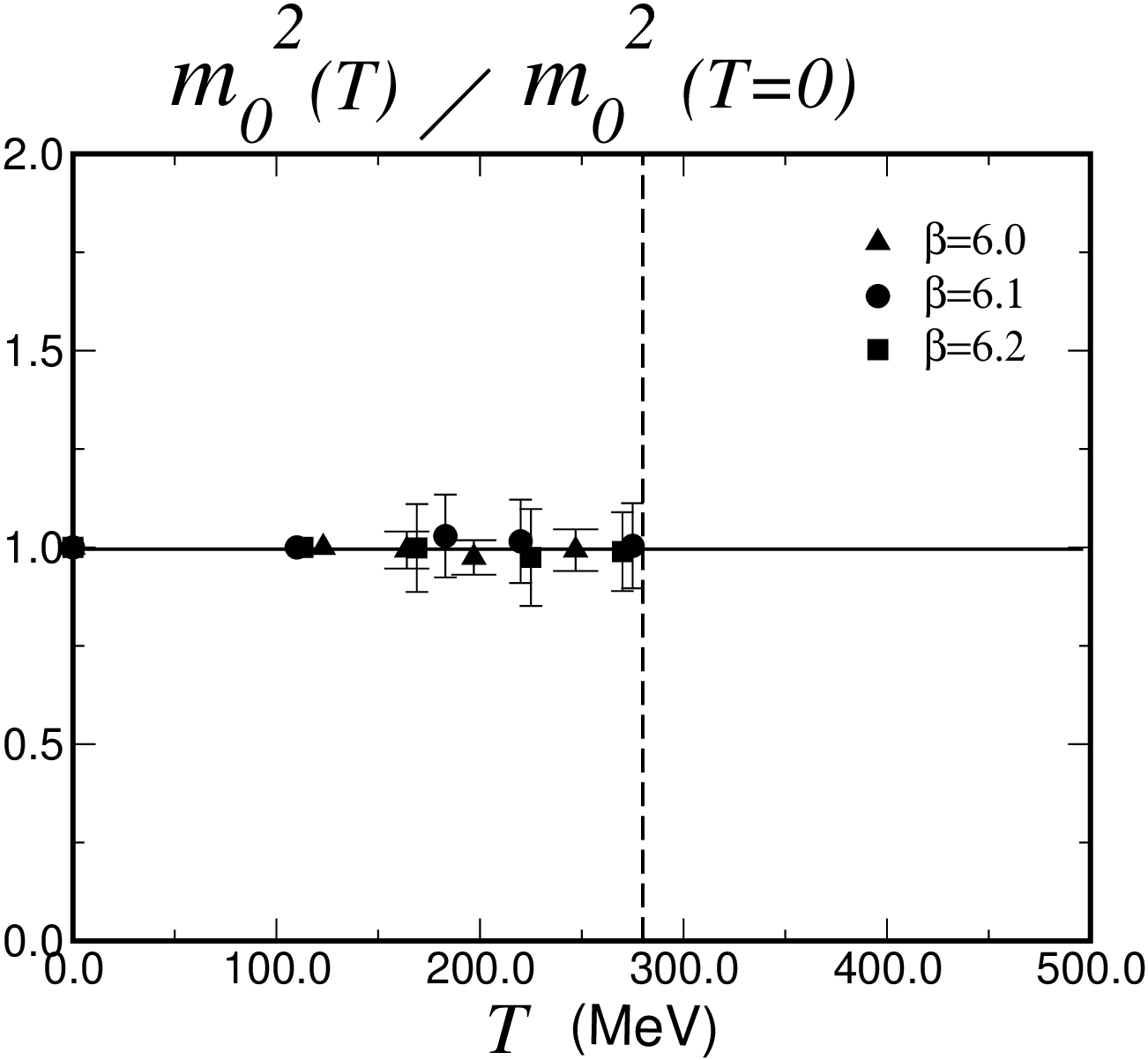}
\caption{The thermal effects on $m_0^2 \equiv \qGq/\qq$ v.s.
the temperature $T$.
This result indicates the same critical behavior 
between $\qq$ and $\qGq$.
}
\end{center}
\label{fig:M0_finite_T}
\end{figure}

We then compare the thermal effects on
$\qGq$ and $\qq$,
because these two condensates 
characterize different aspects of the QCD vacuum.
For this purpose, we plot the thermal effects 
on $m_0^2 \equiv \qGq / \qq$ in figure~\ref{fig:M0_finite_T}.
%
%
%
%
From figure~\ref{fig:M0_finite_T}, we observe that $m_0^2$ is almost 
independent of the temperature, even in the very vicinity of $T_c$.
This very nontrivial result means that 
{\it both of $\qGq$ and $\qq$ 
obey the same critical behavior.}
This may indicate the existence of the universal 
behavior of chiral order parameters near $T_c$,
and would impose constraints on the chiral structure of the QCD vacuum.


\vspace*{-3mm}

In summary, we have studied the thermal effects on $\qGq$ as well 
as $\qq$ 
using SU(3)$_c$ lattice QCD with  KS-fermion at the quenched level. 
While the thermal effects on both the condensates
are very weak below $T_c$, 
we have observed a strong chiral phase transition near $T_c$
as a sudden disappearance of both the condensates.
We have also observed that 
$m_0^2 \equiv \qGq /\qq$ 
is almost independent of the temperature
even in the very vicinity of $T_c$, 
which means
both the condensates obey the same critical behavior.
This nontrivial similarity would impose constraints on 
the chiral structure of the QCD vacuum near $T_c$.
For further studies,
a full QCD lattice calculation is in progress
in order to analyze the dynamical quark effects on the condensates.



\vspace*{-1mm}


\begin{thebibliography}{0}
%
\bibitem{Dosch}     {H.G. Dosch, M. Jamin and S. Narison,
                        {\it Phys. Lett.} {\bf B220}, 251 (1989).}

\bibitem{Dosch2}     {H.G. Dosch and S. Narison,
                        {\it Phys. Lett.} {\bf B417}, 173 (1998).}

\bibitem{Latorre}    {J.I. Latorre, P. Pascual and S. Narison,
                        {\it Z. Phys.} {\bf C34}, 347 (1987).}


\bibitem{Bel}       {V.M. Belyaev and B.L. Ioffe,
                        {\it Sov. Phys. JETP} {\bf 56}, 493 (1982).}
%
\bibitem{K&S}       {M. Kremer and G. Schierholz,
                        {\it Phys. Lett.} {\bf B194}, 283 (1987).}
%
\bibitem{DOIS:qGq}  {T. Doi, N. Ishii, M. Oka and H. Suganuma,
			{\it Phys. Rev.} {\bf D67}, 054504 (2003).}
%
\bibitem{DOIS:T}  {T. Doi, N. Ishii, M. Oka and H. Suganuma,
			{\it Nucl. Phys.} {\bf A721}, 934 (2003);
			{\it Prog. Theor. Phys. Suppl.} {\bf 151}, 161 (2003);
                        in Proc. 
                        ``Quark Confinement and the Hadron Spectrum V''
			(World Sci., 2003) p.381,
                        hep-lat/0212025.}
%
\bibitem{twc:qGq}   {T.W. Chiu and T.H. Hsieh, \ hep-lat/0305016.} 
%
\bibitem{DOIS:T2}  {T. Doi, N. Ishii, M. Oka and H. Suganuma,
			{\it Nucl. Phys.} {\bf B} (Proc. Suppl.)
			in press,
			hep-lat/0309124.}
%
\bibitem{Narison2}  { S. Narison and R. Tarrach,
                        {\it Phys. Lett.} {\bf B125}, 217 (1983).}
%
\end{thebibliography}
\end{document}